\documentclass[twocolumn]{revtex4-1}
\usepackage{amsmath}
\usepackage{graphicx}
\usepackage{amssymb}
\usepackage{color}

\newcommand{\be}{\begin{eqnarray}}
\newcommand{\ee}{\end{eqnarray}}

\newcommand{\p}{\partial}

\newcommand{\nn}{\nonumber}

\newcommand{\diag}{\mathop{\rm diag}}

\newcommand{\kL}{\mathfrak{L}}

\newcommand{\cA}{{\mathcal A}}

\newcommand{\cF}{{\mathcal F}}
\newcommand{\cQ}{{\mathbf F}}

\newcommand{\cD}{{\mathcal D}}
\newcommand{\cG}{{\mathcal G}}
\newcommand{\fG}{{\mathbf G}}
\newcommand{\cW}{{\mathcal W}}

\newcommand{\dchi}{\delta \chi }

\newcommand{\1}{\mspace{1mu}}

\newcommand{\fGa}{\mathbf \Gamma}
\newcommand{\vGa}{\varGamma}

\newcommand{\cR}{\mathcal R}

\newcommand{\fL}{\mathbf L}
\newcommand{\fM}{\mathbf M}
\newcommand{\fN}{\mathbf N}
\newcommand{\fP}{\mathbf P}
\newcommand{\fQ}{\mathbf Q}

\newcommand{\fA}{\mathbf A}
\newcommand{\fB}{\mathbf B}
\newcommand{\fC}{\mathbf C}
\newcommand{\fD}{\mathbf D}

\newcommand{\bet}{\bar{\eth}}

\newcommand{\vOm}{\varOmega }
\newcommand{\cH}{\mathbf A }

\newcommand{\chih}{\hat{\chi}}

\newcommand{\hx}{\hat{x}}

\newcommand{\tx}{\tilde{x}}
\newcommand{\mG}{\mathsf{G}}

\newcommand{\vPsi}{\varPsi}

\begin{document}
\def\intdk{\int\frac{d^4k}{(2\pi)^4}}
\def\sla{\hspace{-0.17cm}\slash}
\hfill


\title{ Unified field theory of basic forces and elementary particles \\ with gravitational origin of gauge symmetry in hyper-spacetime}

\author{Yue-Liang Wu}\email{ylwu@itp.ac.cn}
\affiliation{$^a$International Centre for Theoretical Physics Asia-Pacific (ICTP-AP), Beijing, China \\
$^b$Institute of Theoretical Physics, Chinese Academy of Sciences, Beijing 100190, China\\
$^c$University of Chinese Academy of Sciences (UCAS), Beijing 100049, China }



\maketitle


The theory of general relativity (GR)\cite{GR} has successfully built a relation between the geometry of spacetime and the energy-momentum tensor of matter. Einstein has extensively quested for potential unified models of electromagnetism and gravity at a classical unified field theory and spent the last two decades of his life to the search for a unified field theory.  A scaling gauge symmetry\cite{WG} was proposed by Weyl for an attempt to characterize the electromagnetism. Kaluza and Klein\cite{KK} extended the GR to a five dimensional spacetime for trying to unify the electromagnetism. The relativistic Dirac spinor theory\cite{DE} has remarkably set up a correlation between the dimensions of spacetime and the degrees of freedom of Dirac spinor, that reflects a geometric coherence of spinor field at a profound level. A modern unified field theory is an attempt to find out a theory that can unify the four known basic forces: gravity, electromagnetic, weak and strong. Three of them have well been described by Abelian and non-Abelian Yang-Mills gauge fields\cite{YM}, which lays the foundation of the standard model(SM)\cite{EW,QCD} within the framework of quantum field theory (QFT). The discovery of asymptotic freedom\cite{QCD} indicates a potential unification. The idea of unified theories with enlarged gauge symmetries was initiated from unifying quarks and leptons\cite{PS} and led to the construction of grand unified theories (GUTs) SU(5)\cite{SU5} and SO(10)\cite{SO10} for the electroweak and strong interactions. An enlarged SO(1,13) gauge model\cite{SO14} was proposed to unify SO(1,3) and SO(10).  The unifying groups SO(1,13) and SO(3,11) were also motivated to be a gravity GUT model in four dimensional spacetime\cite{GGUT}.

Inspiring from the relativistic Dirac spinor theory and Einstein theory of GR as well as GUTs, we are going to present a unified field theory for all known basic forces and elementary particles. The construction of the theory is based on the postulate of gauge invariance and coordinate independence described in \cite{YLWU1} for a gravitational gauge field theory within the framework of QFT. 

By treating all spin-like charges of elementary particles on the same footing as a {\it hyper-spin charge} and expressing the degrees of freedom of all elementary particles into a single column vector in a spinor representation of a high-dimensional {\it hyper-spacetime}, we shall be able to establish a coherent relation between the spinor structure of elementary particles and the dimension of a {\it hyper-spacetime}. A minimal unified spinor field is found to be a Majorana-type {\it hyper-spinor field} $\varPsi(\hx)$ in an irreducible spinor representation of a {\it hyper-spin group} SP(1,18)$\cong$SO(1,18), which results in a Minkowski {\it hyper-spacetime} with dimension $D_h=19$.  

For a freely moving massless {\it hyper-spinor field} $ \varPsi(\hx)$ in a Minkowski {\it hyper-spacetime}, we can write down a self-hermitian action via a maximal symmetry 
\be \label{UFTaction1}
I_H = \int [d\hx] \, \frac{1}{2} \bar{\vPsi}(\hx) \varGamma^{\fA} \, \delta_{\fA}^{\;\; \fM} i \partial_{\fM} \vPsi(\hx) \, , 
\ee
with $\hx \equiv x^{\fM}$ and $\fA, \fM =0,1,2,3,5, \cdots, D_h$ ($D_h = 19$). Where $\partial_{\fM}\equiv \partial/\partial x^{\fM}$ is the partial derivative and $\delta_{\fA}^{\;\;\fM} $ the Kronecker symbol. The Latin alphabets $\fA,\cdots$ and those starting from $\fM $ are used to distinguish the vector indices in a non-coordinate spacetime and a coordinate spacetime, respectively.  All Latin indices are raised and lowered by the constant metric matrices, i.e., $\eta^{\fA\fB}=\diag(1,-1,\cdots,-1)$, and $\eta^{\fM\fN} =\diag(1,-1,\cdots,-1)$. 

The structures of the Majorana-type {\it hyper-spinor field} $ \varPsi(\hx)$ and the $\gamma$-matrix $\varGamma^{\fA} \equiv ( \varGamma^a, \varGamma^{A}, \varGamma^{m} )$ ($a=0,1,2,3$, $A = 5,\cdots, 14$, $m = 15,\cdots, D_h$) are found to be 
\be \label{19DMHSF}
\varPsi(\hx) =
\begin{pmatrix}
\Psi(\hx) \\
\Psi^{c}(\hx)\\
\Psi'(\hx) \\
-\Psi^{'c}(\hx)
\end{pmatrix} ; 
\begin{matrix}
& & \varGamma^{a} = \; \sigma_0 \otimes \sigma_0 \otimes I_{32}\otimes \gamma^a\, ,  \\
& & \varGamma^{A} = i \sigma_0 \otimes \sigma_0 \otimes \Gamma^{A}\otimes \gamma_5\, ,  \\
& & \varGamma^{15} = i \sigma_2 \otimes \sigma_3\otimes \gamma^{11} \otimes \gamma_5\, ,\\
& & \varGamma^{16} = i \sigma_1 \otimes \sigma_0\otimes  \gamma^{11} \otimes \gamma_5\, , \\
& & \varGamma^{17} = i \sigma_2 \otimes \sigma_1\otimes \gamma^{11} \otimes \gamma_5\, , \\
& & \varGamma^{18} = i \sigma_2 \otimes \sigma_2\otimes \gamma^{11}\otimes \gamma_5\, , \\
& & \varGamma^{19} = \; \sigma_0 \otimes \sigma_0 \otimes I_{32}\otimes I_{4} \, , 
\end{matrix}
\ee
with $\sigma_i$ (i=1,2,3) the Pauli matrices, $\sigma_0\equiv I_2$, $I_{4}$ and $I_{32}$ the unit matrices. $\gamma^a$ and $\Gamma^A$ are the known $\gamma$-matrices defined in four and ten dimensions, respectively. The action Eq.(\ref{UFTaction1}) possesses a maximal hyper-spin symmetry 
\be
SP(1,D_h -1)\cong SO(1,D_h -1)\, , \quad D_h =19\, ,
\ee
with generators $\varSigma^{\fA\fB} = -  \varSigma^{\fB\fA} = i[\varGamma^{\fA}\, , \varGamma^{\fB}]/4$ and  $\varSigma^{\fA19} = - \varSigma^{19\fA} = i\varGamma^{\fA}/2\, $, $\fA, \fB = 0, 1,2,3,5,\cdots, 18$.

The spinors $\Psi(\hx)$ and $\Psi'(\hx)$ are Dirac-type hyper-spinor fields with their charge conjugated ones $\Psi^{c}(\hx)$ and $\Psi^{'c}(\hx)$ defined in the 14-dimensions. They have a general hyper-spinor structure 
\be \label{SS1}
 \Psi  \equiv \Psi_{W} + \Psi_{E} \equiv \Psi_1 + i \Psi_2 \, ,
\ee
with $\Psi_{W, E} = \frac{1}{2} (1 \mp \gamma_{15}) \Psi $ and $\Psi_i \equiv  \Psi_{W i} + \Psi_{E i}$ ($i=1,2$). $\Psi_{W i}$ and $\Psi_{E i}$ are regarded as a pair of {\it mirror hyper-spinor fields} referred as {\it westward} and {\it eastward } hyper-spinor fields. Each {\it westward} spinor field $\Psi_{W i}$ identifies to a family of quarks and leptons in SM
\be
& & \Psi_{W\1 i}^{T} =  [ (U_i^{r}, U_i^{b}, U_i^{g}, U_i^{w}, D^{r}_{ic}, D^{b}_{ic}, D^{g}_{ic}, D^{w}_{ic}, \nn \\
& & \quad \;  D_i^{r}, D_i^{b}, D_i^{g}, D_i^{w}, -U^{r}_{ic}, -U^{b}_{ic}, -U^{g}_{ic}, -U^{w}_{ic})_L\, ,  \nn \\
& & \quad \; \ (U_i^{r}, U_i^{b}, U_i^{g}, U_i^{w}, D^{r}_{ic}, D^{b}_{ic}, D^{g}_{ic}, D^{w}_{ic}, \nn \\
& & \quad \; D_i^{r},  D_i^{b}, D_i^{g}, D_i^{w}, -U^{r}_{ic}, -U^{b}_{ic}, -U^{g}_{ic}, -U^{w}_{ic})_R ]^T ,
\ee
which is a Majorana-Weyl type hyper-spinor field in 14-dimensions with 64 independent degrees of freedom. $Q_i^{\alpha} = (U_i^{\alpha}, D_i^{\alpha})$ are the Dirac spinors of quarks and leptons with $\alpha= (r,\, b\, , g\, , w)$ representing the trichromatic (red, blue, green) and white colors, and $Q_{i c}^{\alpha} =(U_{i c}^{\alpha},  D_{i c}^{\alpha})$ the conjugated ones $Q_{i\1 c}^{\alpha} = C_4 \bar{Q}_i^T$ with $C_4$ defined in 4D-spacetime. The subscripts $``L"$ and $``R"$ in $Q_{i L,R}^{\alpha}$ denote the left-handed and right-handed Dirac spinor. Each hyper-spinor field $\Psi_{W i}$ or $\Psi_{E i}$ satisfies a Majorana-Weyl type condition defined in the 14-dimensions with a W-parity\cite{YLWU2}, i.e., $\Psi_{W i,E i}^{c} = C_{14} \bar{\Psi}_{W i, E i}^T = \gamma_5 \Psi_{W i, E i}$ and $\gamma_{15} \Psi_{W i, E i} = \mp \Psi_{W i, E i}$ with $\gamma_{15} = \gamma_{11}\otimes\gamma_5$.

The action Eq.(\ref{UFTaction1}) is invariant under the discrete symmetries: charge-conjugation $\mathcal{C}$, parity-inversion  
 $\mathcal{P}$ and time-reversal $\mathcal{T}$ defined in the hyper-spacetime with dimension $D_h =19$. The hyper-spinor field transforms as 
\be
\mathcal{C} \vPsi(\hx ) \mathcal{C}^{-1} & = & C_{19} \bar{\vPsi}^{T} ( \hx )= \vPsi(\hx)\, , \; C_{19}^{-1} \varGamma^{\fA} C_{19} = (\varGamma^{\fA})^{T}, \nn \\
\mathcal{P} \vPsi(\hx ) \mathcal{P}^{-1} & = & P_{19} \vPsi(\tx )\, , \quad P_{19}^{-1} \varGamma^{\fA} P_{19} = ( \varGamma^{\fA})^{ \dagger}, \nonumber \\
 \mathcal{T} \vPsi(\hx ) \mathcal{T}^{-1} & = & T_{19} \vPsi( -\tx ) \, , \quad T_{19}^{-1} \varGamma^{\fA} T_{19} = (\varGamma^{\fA})^{T}\, ,
\ee
with $\tx \equiv (x^0, -x^{1},\cdots, - x^{18}, x^{19}) $, and
\be
C_{19} & = & \varGamma_2\varGamma_0\varGamma_6\varGamma_8\varGamma_{10}\varGamma_{12}\varGamma_{14}  \varGamma_{16}\varGamma_{18} =  - i \sigma_3\otimes \sigma_2 \otimes C_{14}\, , \nn \\
P_{19} & = &  \varGamma_0;\;\;  T_{19}  = i \varGamma_1\varGamma_3 \varGamma_5\varGamma_7 \varGamma_9\varGamma_{11} \varGamma_{13}\varGamma_{15} \varGamma_{17}\gamma_{19}.
\ee
A hyper-spinor field equation is obtained from Eq.(\ref{UFTaction1})
\be
 \varGamma^{\fA} \, \delta_{\fA}^{\;\; \fM} i \partial_{\fM} \vPsi(\hx) = 0\, , \quad \eta^{\fM\fN}\p_{\fM}\p_{\fN} \vPsi(\hx) = 0\, , 
\ee
with $\eta^{\fM\fN} = \diag(1, -1, \cdots , -1)$. Such an equation leads to a generalized relativistic quantum theory for a hyper-spinor field in a globally flat hyper-spacetime.

In a Minkowski hyper-spacetime, the hyper-spin symmetry SP(1,$D_h$-1) and Lorentz symmetry SO(1,$D_h$-1) have to incorporate coherently. Based on a gauge principle, a fundamental interaction is postulated to be governed by taking SP(1,$D_h$-1) as a local hyper-spin gauge symmetry. To realize that, it is essential to introduce a {\it bicovaraint vector field} $\chih_{\fA}^{\;\, \fM}(\hx)$ and a {\it hyper-spin gauge field} $\cA_{\fM}(\hx) \equiv \cA_{\fM}^{\; \fA\fB}(\hx)\, \frac{1}{2}\varSigma_{\fA\fB}$. By requiring the theory be invariant under both global and local conformal scaling transformations, a {\it scaling scalar field} $\phi(\hx)$ is needed. 

With the above consideration, the action Eq.(\ref{UFTaction1}) can be extended to be a gauge invariant one
\be \label{UFTaction2}
I_H =  \int [d\hx] \, \phi^{D_h-4} \chi(\hx) \1 \frac{1}{2} \bar{\vPsi}(\hx) \varGamma^{\fA} \, \chih_{\fA}^{\;\; \fM}(\hx) i \cD_{\fM} \vPsi(\hx) ,
\ee
with $i\cD_{\fM} \equiv i\partial_{\fM} + \cA_{\fM}$ a covariant derivative and $\chi(\hx)$ an inverse of determinant $\chih(\hx) = \det \chih_{\fA}^{\; \; \fM}(\hx)$. Obviously, Eq.(\ref{UFTaction2}) has global and local conformal scaling symmetries S(1) and SG(1), respectively, under the transformations 
\be
& & x^{'\fM} = \lambda^{-1} \1 x^{\fM}\, ;\;\;  \, \cA'_{\fM}(\hx') =  \lambda\1 \cA_{\fM}(\hx) \, , \nn \\
& & \vPsi'(\hx') = \lambda^{3/2}\1 \vPsi(\hx)\, ;\; \phi'(\hx') = \lambda \phi(\hx) \, .  \nn \\
&&  \hat{\chi}_{\fA}^{'\; \fM}(\hx) = \xi(\hx) \hat{\chi}_{\fA}^{\;\; \fM}(\hx)  \, ; \; \chi'(\hx) = \xi^{-D_h}(\hx) \chi(\hx)\, , \nn \\
& & \vPsi'(\hx) = \xi^{3/2} (\hx) \vPsi(\hx) \, ;\; \phi'(\hx) = \xi(\hx)\phi(\hx)\, .
\ee

The action Eq.(\ref{UFTaction2}) possesses a maximal symmetry
\be \label{MS}
G = P(1,D_h\mbox{-}1)\times S(1) \times SP(1,D_h\mbox{-}1)\times SG(1)\, ,
\ee
with P(1,$D_h$-1) = SO(1,$D_h$-1)$\ltimes P^{1,D_h\mbox{-}1}$ a Poincar\'e symmetry group in a Minkowski hyper-spacetime.
 
From Eq.(\ref{UFTaction2}), we can derive an equation of motion
\be \label{EMSF}
& & \varGamma^{\fA} \hat{\chi}_{\fA}^{\;\; \fM}(\hx) i ( {\mathcal D} _{\fM} + {\mathsf V} _{\fM}(\hx) ) \vPsi(\hx) =   0 \, , 
\ee
with ${\mathsf V} _{\fM}$ viewed as an {\it induced-gauge field} 
\be
{\mathsf V}_{\fM} (\hx) & = & \frac{1}{2} \partial_{\fM}\ln (\chi \phi^{D_h-3})  -\frac{1}{2}\chih_{\fB}^{\;\; \fN}\cD_{\fN}\chi_{\fM}^{\;\; \fB} \, , \nn \\
\cD_{\fM}\chi_{\fN}^{\;\; \fA} & = & (\, \partial_{\fM} + \p_{\fM} \ln\phi\, ) \chi_{\fN}^{\;\; \fA} + \cA_{\fM\, \fB}^{\fA}  \chi_{\fN}^{\;\;\fB} \, ,   
\ee
which keeps Eq.(\ref{EMSF}) be conformal scaling gauge covariance. $\chi_{\fM}^{\;\; \fA}(\hx)$ defines a dual bicovaraint vector field
\be \label{Dual}
\chi_{\fM}^{\;\; \fA} (\hx)\, \chih^{\;\; \fM}_{\fB}(\hx)   =  \eta^{\;\; \fA}_{\fB}\, , \;\, \chi_{\fM}^{\;\; \fA}(\hx) \chih^{\;\;\fN}_{\fA}(\hx) = \eta_{\fM}^{\;\;\fN} \, .
\ee
$\chi_{\fM}^{\;\; \fA}(\hx)$ characterizes a gravitational interaction in the hyper-spacetime and is referred as a {\it hyper-gravifield}.  

Eq.(\ref{EMSF}) realizes a gravitational relativistic quantum theory of a hyper-spinor field in the hyper-spacetime. A gauge invariant quadratic form is found to be 
\be \label{EMSF2}
& & \chih^{\fM\fN} (\nabla_{\fM} +  {\mathsf V} _{\fM} ) ( {\mathcal D} _{\fN} + {\mathsf V} _{\fN} )  \vPsi 
= \varSigma^{\fA\fB} \chih_{\fA}^{\;\; \fM} \chih_{\fB}^{\;\; \fN} \nn \\ 
& & \quad \cdot [\, \cF_{\fM\fN} + i {\cal V}_{\fM\fN} - {\cal G}_{\fM\fN}^{\fC} \chih_{\fC}^{\;\; \fP} i ( {\mathcal D} _{\fP} + {\mathsf V} _{\fP} )\, ]  \vPsi \, , 
 \ee
with $\nabla_{\fM} \cD_{\fN} \equiv \cD_{\fM}\cD_{\fN} + \fGa_{\fM\fN}^{\fP} \cD_{\fP}$. $\cF_{\fM\fN}(\hx)$, ${\cal G}_{\fM\fN}^{\fA}$ and ${\cal V}_{\fM\fN}$  define the field strengths of the {\it hyper-spin gauge field} $\cA_{\fM}(\hx)$, the {\it hyper-gravifield} $\chi_{\fM}^{\; \fA}(\hx)$ and the induced-gauge field  ${\mathsf V}_{\fM}$, respectively, 
 \be \label{STGF}
 \cF_{\fM\fN} & = & \partial_{\fM} {\cal A}_{\fN}- \partial_{\fN} {\cal A}_{\fM} - i [{\cal A}_{\fM},  {\cal A}_{\fN} ] \equiv\cF_{\fM\fN}^{\fA\fB}\frac{1}{2}\varSigma_{\fA\fB}  \, , \nn \\
 {\cal G}_{\fM\fN}^{\fA} & = & \cD_{\fM} \chi_{\fN}^{\;\; \fA} - \cD_{\fN} \chi_{\fM}^{\;\; \fA};\;  \chih^{\fM\fN} =  \chih_{\fA}^{\;\; \fM} \chih_{\fB}^{\;\; \fN} \eta^{\fA\fB}, \nn \\
 {\cal V}_{\fM\fN} & = &  ( \partial_{\fM}  {\mathsf V} _{\fN} - \partial_{\fN}  {\mathsf V} _{\fM});  \;\; \fGa_{\fM\fN}^{\fP}  =  \chih_{\fA}^{\;\; \fP}  \cD_{\fM}\chi_{\fN}^{\;\;\fA} \, .
\ee

Let us now apply a framework of gravitational gauge field theory and QFT shown in refs.\cite{YLWU1,YLWU2}  to build a unified field theory of all known basic forces and elementary particles. We shall first define, from a dual basis $(\{\p_{\fM}\}, \{dx^{\fM}\})$ of a coordinate system, a non-coordinate basis via the dual condition
\be \label{eth}
& & \langle \delta \chi^{\fA}, \eth_{\fB}\rangle = \chi_{\fM}^{\;\; \fA}(\hx)  \chih_{\fB}^{\;\; \fN} (\hx)  \langle dx^{\fM} , \partial_{\fN} \rangle = \eta_{\fB}^{\;\, \fA}\, , \nn \\
& &  \delta\chi^{\fA} \equiv \chi_{\fM}^{\;\; \fA} (\hx) dx^{\fM}\, , \quad  \eth_{\fA} \equiv  \chih_{\fA}^{\;\, \fM}(\hx) \partial_{\fM} \, .
\ee
Such a dual basis $(\{\eth_{\fA}\}, \{\delta \chi^{\fA}\})$ is referred as a dual {\it hyper-gravifield basis} and its spanned non-coordinate hyper-spacetime is called as a {\it hyper-gravifield spacetime}.  Note that such a hyper-gravifield basis is no longer commutative and satisfies a non-commutation relation 
\be
& &   [ \eth_{\fA} ,\; \eth_{\fB}] =  f_{\fA\fB}^{\; \fC}\,  \eth_{\fC} \, ; \; \;  f_{\fA\fB}^{\; \fC} \equiv - \chih_{\fA}^{\;\; \fM} \chih_{\fB}^{\;\; \fN} \mG_{\fM\fN}^{\; \fC}, 
\ee 
with $\mG_{\fM\fN}^{\; \fC} \equiv\partial_{\fM}\chi_{\fN}^{\;\; \fC} - \partial_{\fN}\chi_{\fM}^{\;\; \fC}$. It indicates that the locally flat hyper-gravifield spacetime is associated with a noncommutative geometry. Such a spacetime structure forms a {\it biframe hyper-spacetime}. A biframe spacetime may geomatrically be treated in parallel\cite{NCG}. 

With a projection of hyper-gravifield, we are able to construct a gauge invariant and coordinate independent unified field theory in the hyper-gravifield spacetime, 
\be
\label{UFTaction}
I_H & \equiv & \int [\dchi]\; \kL = \int [\dchi]\, \phi^{D_h-4}\{\,  \frac{1}{2}\bar{\vPsi}\vGa^{\fC}i\cD_{\fC} \vPsi  \nonumber \\
& - & \frac{1}{4}\, [\, g_h^{-2}\tilde{\eta}^{\fC\fD\1 \fC'\fD'}_{\fA\fB\1 \fA'\fB' } \cF_{\fC\fD}^{\fA\fB}\cF_{\fC'\fD'}^{ \fA'\fB'} + {\cal W}_{\fC\fD} {\cal W}^{\fC\fD} \, ] \nn \\
& + & \alpha_E \phi^2 [ \frac{1}{4}  \, \tilde{\eta}^{\fC\fD\fC'\fD'}_{\fA\fA'} \cG_{\fC\fD}^{\fA}\cG_{\fC'\fD'}^{\fA'} - \eta_{\fA}^{\fC}\eta_{\fB}^{\fD}\cF_{\fC\fD}^{\fA\fB} ] \nn \\
& + &  \frac{1}{2}\eta^{\fC\fD}\bet_{\fC} \phi \bet_{\fD}\phi  - \beta_E\1 \phi^4  \, \} \, , 
\ee
with $g_h$, $\alpha_E$ and $\beta_E$ the coupling constants. Where $\cA_{\fC}^{\fA\fB} \equiv \chih_{\fC}^{\;\, \fM} \cA_{\fM}^{\fA\fB}$, $W_{\fC} \equiv \chih_{\fC}^{\;\, \fM} W_{\fM}$, $i\cD_{\fC} \equiv i \eth_{\fC}  + \cA_{\fC}$ and $\bet_{\fC} \phi  \equiv (\eth_{\fC} - g_w W_{\fC}) \phi$. $W_{\fM}$ is a Weyl gauge field\cite{WG} introduced to characterize a conformal scaling gauge invariant dynamics of the {\it scaling scalar field} $\phi$. The field strengths in the hyper-gravifield spacetime are given by:  $\cF_{\fC\fD}^{\fA\fB} \equiv \cF_{\fM\fN}^{\fA\fB}  \chih^{\;\fM}_{\fC} \chih^{\;\fN}_{\fD} $, $\cG_{\fC\fD}^{\fA}\equiv \cG_{\fM\fN}^{\fA}\chih^{\;\fM}_{\fC} \chih^{\;\fN}_{\fD}$ and $\cW_{\fC\fD}  \equiv \cW_{\fM\fN}\chih^{\fM}_{\fC} \chih^{\fN}_{\fD}$ with $\cW_{\fM\fN} = \p_{\fM}W_{\fN} - \p_{\fN}W_{\fM}$.

The tensor $\tilde{\eta}^{\fC\fD\1 \fC'\fD'}_{ \fA\fB\1\fA'\fB'} $ takes the following structure to achieve a general conformal scaling gauge invariance,
\be \label{tensor}
\tilde{\eta}^{\fC\fD\1 \fC'\fD'}_{ \fA\fB\1\fA'\fB'} & \equiv & \frac{1}{4}\1 \{\, [\, \eta^{\fC\fC'} \eta_{\fA\fA'} (\eta^{\fD\fD'} \eta_{\fB\fB'} - 2  \eta^{\fD}_{\fB'} \eta^{\fD'}_{\fB}) \nn \\
& + & \eta^{(\fC,\fC'\leftrightarrow\fD, \fD' )}\, ]+  \eta_{(\fA,\fA'\leftrightarrow \fB,\fB' ) }  \, \} \nn \\
&  + & \frac{1}{4}\alpha_W\, \{\, [\, (\eta^{\fC}_{\fA'} \eta^{\fC'}_{\fA} - 2\eta^{\fC\fC'} \eta_{\fA\fA'}  )  \eta^{\fD}_{\fB'}  \eta^{\fD'}_{\fB}  \nn \\
& + &  \eta^{(\fC,\fC'\leftrightarrow\fD, \fD' )} \, ] +  \eta_{(\fA,\fA'\leftrightarrow \fB,\fB' ) }  \,  \}    \nn \\
& + & \frac{1}{2}\beta_W\, \{\, [\, (\eta_{\fA\fA'}  \eta^{\fC\fC'} - \eta^{\fC'}_{\fA}\eta^{\fC}_{\fA'}) \eta^{\fD}_{\fB} \eta^{\fD'}_{\fB'}\nn \\
& + & \eta^{(\fC,\fC'\leftrightarrow\fD, \fD' )}\, ]+  \eta_{(\fA,\fA'\leftrightarrow \fB,\fB' )} \, \}  \, ,
\ee
with two coupling constants $\alpha_W$ and $\beta_{W}$. The tensor factor $\tilde{\eta}^{\fC\fD\fC'\fD'}_{\fA\fA'}$  acquires the following structure to arrive at a general hyper-spin gauge symmetry 
\be  \label{tensorM}
 \tilde{\eta}^{\fC\fD\fC'\fD'}_{\fA\fA'}  & \equiv &    \eta^{\fC\fC'} \eta^{\fD\fD'} \eta_{\fA\fA'}  
+  \eta^{\fC\fC'} ( \eta_{\fA'}^{\fD} \eta_{\fA}^{\fD'}  -  2\eta_{\fA}^{\fD} \eta_{\fA'}^{\fD'}  ) \nn \\
& + & \eta^{\fD\fD'} ( \eta_{\fA'}^{\fC} \eta_{\fA}^{\fC'} -2 \eta_{\fA}^{\fC} \eta_{\fA'}^{\fC'} ) \, .
\ee

When projecting Eq.(\ref{UFTaction}) into the hyper-spacetime via the hyper-gravifield $\chi_{\fM}^{\;\fA}$, we arrive at a gauge invariant unified field theory within the framework of QFT,
\be \label{UFTactionQFT}
 & & I_H   \equiv  \int [d\hx]  \chi\,  \{\, \frac{1}{2} \bar{\vPsi} \varGamma^{\fA} \chih_{\fA}^{\; \fM} ( i \p_{\fM} + g_h\1 \cA_{\fM} )  \vPsi  
  \nn \\
 & & + \phi^{D_h-4}[ - \frac{1}{4} (\tilde{\chi}^{\fM\fN\1 \fM'\fN'}_{\fA\fB\1 \fA'\fB' } \cF_{\fM\fN}^{\fA\fB}\cF_{\fM'\fN'}^{ \fA'\fB'} +  {\cal W}_{\fM\fN} {\cal W}^{\fM\fN} )  \nn \\
&  & + \alpha_E \phi^2 \frac{1}{4} \tilde{\chi}^{\fM\fN\fM'\fN'}_{\fA\fA'} \fG_{\fM\fN}^{\fA}\fG_{\fM'\fN'}^{\fA'}  +  \frac{1}{2}\hat{\chi}^{\fM\fN} d_{\fM} \phi d_{\fN}\phi   \nn \\
&  & - \beta_E\phi^4 ] \}  +   2\alpha_Eg_h\p_{\fM}(\chi \phi^{D_h-2} \cA_{\fN}^{\fN\fM}),
\ee
with the field strength and tensors defined as
\be
& & \fG_{\fM\fN}^{\fA} =  \hat{\p}_{\fM}\chi_{\fN}^{\; \fA}  -   \hat{\p}_{\fN}\chi_{\fM}^{\; \fA}; \;\; \hat{\p}_{\fM}\equiv \p_{\fM} + \p_{\fM} \ln\phi,  \nn \\
& & \tilde{\chi}^{\fM\fN\1 \fM'\fN'}_{\fA\fB\1 \fA'\fB' }\equiv \chih_{\fC}^{\;\fM}\chih_{\fD}^{\;\fN} \chih_{\fC'}^{\;\fM'}\chih_{\fD'}^{\;\fN'} \tilde{\eta}^{\fC\fD\1 \fC'\fD'}_{\fA\fB\1 \fA'\fB' }\, , \nn \\
& & \tilde{\chi}^{\fM\fN\1 \fM'\fN'}_{\fA\fA'}\equiv \chih_{\fC}^{\;\fM}\chih_{\fD}^{\;\fN} \chih_{\fC'}^{\;\fM'}\chih_{\fD'}^{\;\fN'} \tilde{\eta}^{\fC\fD\1 \fC'\fD'}_{\fA\fA' }\, ,
\ee
and $d_{\fM} \phi =    (\p_{\fM} - g_wW_{\fM})\phi$. A redefinition for the hyper-spinor field $\vPsi\to \vPsi/\phi^{(D_h-4)/2}$ and hyper-spin gauge field $\cA_{\fM}\to g_h\cA_{\fM}$ has been made. The last term in Eq.(\ref{UFTactionQFT}) reflects a surface effect with $\cA_{\fN}^{\fN\fM} \equiv \chih_{\fA}^{\, \fN}\chih_{\fB}^{\, \fM}\cA_{\fN}^{\fA\fB}$.

Note that it is the tensor structure $\tilde{\eta}^{\fC\fD\fC'\fD'}_{\fA\fA'}$ that leads gravitational mass-like terms of the hyper-spin gauge field $\cA_{\fM}^{\fA\fB}$ to be absent from the action Eq.(\ref{UFTactionQFT}). The field strength $\fG_{\fM\fN}^{\fA}$ that describes a gauge gravitational interaction corroborates a {\it gauge gravity correspondence}.

Eq.(\ref{UFTactionQFT}) has in general an enlarged local symmetry,
\be
G_{MS} = GL(D_h, R) \times \mbox{SP(1},D_h\mbox{-1)} \times \mbox{SG(1)}, 
\ee
with GL($D_h$,R) viewed as a hidden general linear group symmetry, which lays the foundation of Einstein's GR in four dimensions. Such a symmetry emerges due to the fact that all gauge field strengths in Eq.(\ref{UFTactionQFT}) are antisymmetry tensors in the hyper-spacetime though the action involves no interactions of a connection $\vGa_{\fM\fN}^{\fP}$ that characterizes the local symmetry GL($D_h$,R). It is actually a consequence of the postulate of gauge invariance and coordinate independence, which becomes a more general principle that allows us to choose a Minkowski hyper-spacetime instead of a Riemannian hyper-spacetime as a base spacetime. The action Eq.(\ref{UFTactionQFT}) is deemed to have a maximal global and local symmetry given in Eq.(\ref{MS}). 

The hyper-gravifield $\chi_{\fM}^{\; \fA}$ is introduced as an accompaniment of the hyper-spin gauge field $\cA_{\fM}^{\fA\fB} $ to ensure the hyper-spin gauge symmetry, it is conceivable to associate $\chi_{\fM}^{\; \fA}$ with the hyper-spin gauge symmetry. Decompose $\cA_{\fM}^{\fA\fB}$ into two parts $\vOm_{\fM}^{\fA\fB} $ and $\cH_{\fM}^{\fA\fB}$, so that $\vOm_{\fM}^{\fA\fB}$ presents an inhomogeneous gauge transformation and $\cH_{\fM}^{\fA\fB}$ transforms homogeneously. Explicitly, we have
\be \label{GOGS}
\vOm_{\fM}^{\fA\fB} & = & \frac{1}{2}[ \hat{\chi}^{\fA\fN} \mG_{\fM\fN}^{\fB} - \hat{\chi}^{\fB\fN} \mG_{\fM\fN}^{\fA} -  \hat{\chi}^{\fA\fP}  \hat{\chi}^{\fB\fQ}  \mG_{\fP\fQ}^{\fC} \chi_{\fM \fC } ], \nn \\
 \mG_{\fM\fN}^{\fA} & \equiv & \p_{\fM} \chi_{\fN}^{\; \fA} -  \p_{\fN} \chi_{\fM}^{\;\fA} \, , \quad \cA_{\fM}^{\fA\fB}\equiv \vOm_{\fM}^{\fA\fB} + \cH_{\fM}^{\fA\fB}.
\ee 
$\vOm_{\fM}^{\fA\fB}$ does transform as a gauge field in the adjoint representation of SP(1,$D_h$-1) when $\chi_{\fM}^{\; \fA}$ transforms as a vector,
\be
& &  \vOm_{\fM}^{'\fA\fB} = \Lambda^{\fA}_{\, \fC}  \Lambda^{\fB}_{\, \fD} \vOm_{\fM}^{\fC\fD}  + \frac{i}{2} ( \Lambda^{\fA}_{\, \fC}  \partial_{\fM}  \Lambda^{\fB\fC} - \Lambda^{\fB}_{\, \fC}  \partial_{\fM}  \Lambda^{\fA\fC}), \nn \\
& & \chi_{\fM}^{'\; \fA}(\hx) =   \Lambda^{\fA}_{\;\; \fC}(\hx) \chi_{\fM}^{\; \fC}(\hx)\, , \;  
\Lambda^{\fA}_{\; \fC}(\hx)  \in \mbox{SP(1,$D_h$-1)}\, ,
\ee
$\vOm_{\fM}^{\fA\fB}$ is referred as a {\it hyper-spin gravigauge field} and  $\cH_{\fM}^{\fA\fB}$ as a {\it hyper-spin homogauge field} with field strengths,
\be \label{FS}
\cR_{\fM\fN}^{\fA\fB} & = & \partial_{\fM} \vOm_{\fN}^{\fA\fB} - \partial_{\fN} \vOm_{\fM}^{\fA\fB} + \vOm_{\fM \fC}^{\fA} \vOm_{\fN}^{\fC \fB} -  \vOm_{\fN \fC}^{\fA} \vOm_{\fM}^{\fC \fB} \nn \\  
\cQ_{\fM\fN}^{\fA\fB} & = &  \cD_{\fM} \cH_{\fN}^{\fA\fB} - \cD_{\fN} \cH_{\fM}^{\fA\fB} +  \cH_{\fM \fC}^{\fA} \cH_{\fN}^{\fC \fB} -  \cH_{\fN \fC}^{\fA} \cH_{\fM}^{\fC \fB} \nn \\
\cF_{\fM\fN}^{\fA\fB} & = & \cR_{\fM\fN}^{\fA\fB} + \cQ_{\fM\fN}^{\fA\fB}\, , 
\ee
with $\cD_{\fM} \cH_{\fN}^{\fA\fB}  =  \partial_{\fM}  \cH_{\fN}^{\fA\fB}  +  \vOm_{\fM \fC}^{\fA} \cH_{\fN}^{\fC \fB} - \vOm_{\fM \fC}^{\fB} \cH_{\fN}^{\fC \fA}$.

As the hyper-spin gauge symmetry is characterized by the gauge-type hyper-gravified $\chi_{\fM}^{\; \fA}$ that behaves as a Goldstone-like boson, we come to the statement that the hyper-spin gauge symmetry has a {\it gravitational origin}. 

Such a gravitational origin of gauge symmetry Eq.(\ref{GOGS}) enables us to yield a general relation for the field strength,  
\be \label{relation}
\cF_{\fM\fN}^{\fA\fB} =  (\cR_{\fM\fN}^{\fP\fQ} +  \cQ_{\fM\fN}^{\fP\fQ} )\chi^{\;\fA}_{\fP} \chi^{\;\fB}_{\fQ} \, , 
\ee
where $\cR_{\fM\fN}^{\fP\fQ}\equiv\cR_{\fM\fN\fQ'}^{\fP}\chih^{\fQ'\fQ} $ defines a Riemann tensor of a {\it hyper-spacetime gravigauge field} $\vGa_{\fM\fQ}^{\fP}\equiv \chih_{\fA}^{\;\fP}(\p_{\fM}\chi_{\fQ}^{\;\fA} + \vOm_{\fM\fB}^{\fA}\chi_{\fQ}^{\fB})$, and $\cQ_{\fM\fN}^{\fP\fQ}$ is a field strength of a {\it hyper-spacetime homogauge field} $\cH_{\fM}^{\fP\fQ} \equiv  \chih_{\fA}^{\fP} \chih_{\fB}^{\fQ} \cH_{\fM}^{\fA\fB}$, i.e.,
\be \label{HSGF}
& & \cQ_{\fM\fN}^{\fP\fQ} =  \nabla_{\fM} \cH_{\fN}^{\fP\fQ} - \nabla_{\fN} \cH_{\fM}^{\fP\fQ} + \cH_{\fM\fL}^{\fP} \cH_{\fN}^{\fL\fQ} - \cH_{\fN\fL}^{\fP} \cH_{\fM}^{\fL\fQ} \nn \\
& & \nabla_{\fM} \cH_{\fN}^{\fP\fQ} = \p_{\fM} \cH_{\fN}^{\fP\fQ} + \vGa_{\fM\fL}^{\fP} \cH_{\fN}^{\fL\fQ} + \vGa_{\fM\fL}^{\fQ}  \cH_{\fN}^{\fP\fL},  \nn \\
& & \cR_{\fM\fN\fQ}^{\fP} = \partial_{\fM} \vGa_{\fN\fQ}^{\fP} - \partial_{\fN} \vGa_{\fM\fQ}^{\fP}  + \vGa_{\fM\fL}^{\fP} \vGa_{\fN\fQ}^{\fL}  - \vGa_{\fN\fL}^{\fP} \vGa_{\fM\fQ}^{\fL}, \nn \\
& & \vGa_{\fM\fQ}^{\fP} =  \frac{1}{2} \hat{\chi}^{\fP\fL} (\, \p_{\fM} \chi_{\fQ \fL} + \p_{\fQ} \chi_{\fM \fL} - \p_{\fL}\chi_{\fM\fQ} \, )\, ,
 \ee
with $\vGa_{\fM\fQ}^{\fP}$ as a Christoffel symbol characterized by the hyper-gravimetric field $\chi_{\fM\fN}= \chi_{\fM}^{\; \fA}\chi_{\fN}^{\; \fB} \eta_{\fA\fB}$. 

The relations in Eqs.(\ref{relation}-\ref{HSGF}) indicate a {\it gravity geometry correspondence}, which enables us to rewrite Eq.(\ref{UFTactionQFT}) into an equivalent action in a hidden gauge formalism,
\be \label{UFTactionGG}
& & I_H =   \int d\hx \chi  \{ \frac{1}{2}\bar{\vPsi}\vGa^{\fM} [ i \p_{\fM}  +  (\varXi_{\fM}^{\fP\fQ} + g_h\cH_{\fM}^{\fP\fQ} ) \frac{1}{2}\varSigma_{\fP\fQ} ] \vPsi  \nonumber \\
& &   +\phi^{D_h-4}   \{ -\frac{1}{4} (\, \tilde{\chi}^{\fM\fN\1 \fM'\fN'}_{\fP\fQ\1\fP'\fQ'}  \cQ_{\fM\fN}^{\fP\fQ} \cQ_{\fM'\fN'}^{\fP'\fQ'}  + {\cal W}_{\fM\fN} {\cal W}^{\fM\fN} \, ) \nonumber \\
& &  + \alpha_E  [ \,\phi^2\cR - (D_h-1)(D_h-2)\p_{\fM}\phi\p^{\fM}\phi  \, ]  - \beta_E\1 \phi^4   \nn \\
& & + \frac{1}{2}\hat{\chi}^{\fM\fN} d_{\fM} \phi d_{\fN}\phi  \} \} +  2\alpha_Eg_h\p_{\fM}(\chi \phi^{D_h-2} \cH_{\fN}^{\fN\fM}) ,
\ee
with $\vGa^{\fM}\equiv\chih_{\fA}^{\fM}\vGa^{\fA}$. Where $\varXi_{\fM}^{\fP\fQ}$ defines a pure gauge-type field $\varXi_{\fM}^{\fP\fQ}\equiv \frac{1}{2} ( \chih^{\;\fP}_{\fC} \p_{\fM}\chih^{\fQ\fC} - \chih^{\;\fQ}_{\fC} \p_{\fM}\chih^{\fP\fC} )$, and $\cR$ is a Ricci scalar tensor $\cR \equiv -\cR_{\fM\fN\fQ}^{\fP}\eta_{\fP}^{\fM}\chih^{\fN\fQ}$. 

The tensor field $\tilde{\chi}^{\fM\fN\1 \fM'\fN'}_{\fP\fQ\1\fP'\fQ'}$ is defined via Eq.(\ref{tensor}) as 
\be
\tilde{\chi}^{\fM\fN\1 \fM'\fN'}_{\fP\fQ\1\fP'\fQ'}  \equiv  \chi^{\fA}_{\fP} \chi^{\fB}_{\fQ} \chi^{\fA'}_{\fP'} \chi^{\fB'}_{\fQ'} \chih_{\fC}^{\fM} \chih_{\fD}^{\fN} \chih_{\fC'}^{\fM'}\chih_{\fD'}^{\fN'} \tilde{\eta}^{\fC\fD\1 \fC'\fD'}_{ \fA\fB\1\fA'\fB'}. 
\ee
It can be verified that it is the structure Eq.(\ref{tensor}) that eliminates high derivative terms of Riemann and Ricci tensors in the action Eq.(\ref{UFTactionGG}). The gravitational interaction is governed solely by the Einstein-Hilbert action term, which affirms a {\it gravity geometry correspondence}.

An equivalence of the actions Eqs.(\ref{UFTactionQFT}) and (\ref{UFTactionGG}) is realized more explicitly by choosing a gauge fixing of SP(1,$D_h$-1) to a unitary gauge that makes a symmetric hyper-gravifield $\chi_{\fM\fA} = \chi_{\fA\fM}$, so that $\chi_{\fM\fN} = (\chi_{\fM\fA})^2$. Such an equivalence reveals a {\it gauge geometry duality}.   

In conclusion, we have constructed a unified field theory for all known basic forces and elementary particles based on the postulate of gauge invariance and coordinate independence. We would like to point out that such a postulate of gauge invariance and coordinate independence is more general and fundament than that of general coordinate invariance proposed by Einstein since the laws of nature should be independent of any choice of coordinates. Such a unified field theory has equivalently been formulated in a non-coordinate hyper-gravifield spacetime shown in Eq.(\ref{UFTaction}) and in a coordinate Minkowski hyper-spacetime given in Eq.(\ref{UFTactionQFT}) as well as in a hidden gauge formalism presented in Eq.(\ref{UFTactionGG}). As their equivalence reveals a gravitational gauge geometry duality, the gravitational force characterized by a gauge field strength of hyper-gravifield is completely dual to the one described solely by an Einstein-Hilbert action. Such a unified field theory is built from a bottom-up approach and demands a hyper-spacetime with a minimal dimension $D_h=19$ and a hyper-spin gauge symmetry SP(1,18) to unify all basic forces and elementary particles of quarks and leptons, it also predicts the existence of bulk mirror quarks and leptons as well as bulk vector-like hyper-spinors from which a dark matter candidate is expected to be identified. As such a SP(1,18) gauge symmetry is resulted by equivalently treating all spin-like charges of quarks and leptons, i.e., helicity spin charge, boost spin charge, chirality spin charge, electric spin charge, isometric spin charge, color spin charge and family spin charge, to be as a hyper-spin charge, it is expected to reproduce naturally all known basic forces via a dynamical and geometric symmetry breaking mechanism. For instance, a dynamical and geometric symmetry breaking mechanism may lead to a naive pattern as follows,
\be
SP(1,18) & \to & SP(1,17)\to SP(1,3)\times SO(10)\times SO(4)\nn \\
 &\to & SP(1,3)\times SU(4)\times SU_L(2)\times SU_R(2) \nn \\
 & \to & SP(1,3) \times SU(3)\times SU(2)\times U(1).
\ee

Finally,  we would like to make a remark that there are a large number of articles in literature discussing extra dimensional theories with various compactification patterns and models, the well-known theories include the 10-dimensional string theory and 11-dimensional M-theory. Unlike those theories, all dimensions in the present unified field theory have a physical origin due to their coherent relations to the basic quantum numbers of quarks and leptons. Such a correlation shall enable us to make appropriate dynamical and geometrical evolution and symmetry breaking mechanism to reduce a nineteen dimensional hyper-spacetime to a real four dimensional spacetime. As the non-coordinate hyper-gravifield spacetime is a dynamically generated one, it requires us to develop in general a nonperturbative approach since their evolution is governed by a highly nonlinear dynamics of hyper-gravifield and hype-spin gauge field, which is analogous to a low energy dynamics of QCD.

\centerline{{\bf Acknowledgement}}

This work was supported in part by the National Science Foundation of China (NSFC) under Grants No. 11690022 \& 11475237, and by the Strategic Priority Research Program of the Chinese Academy of Sciences (CAS), Grant No. XDB23030100, and by the CAS Center for Excellence in Particle Physics (CCEPP).

\end{document}